\newcommand{\be}{\begin{eqnarray}}
\newcommand{\ee}{\end{eqnarray}}
\def\llangle{\left \langle}
\def\rrangle{\right \rangle}
\def\MeV{{\rm MeV}}
\def\GeV{{\rm GeV}}
\def\mmntm{{\bf p}}
\def\pstn{{\bf r}}
\def\fm{{\rm fm}}
\def\lsim{\mathrel{\rlap{\lower3pt\hbox{\hskip1pt$\sim$}}
     \raise1pt\hbox{$<$}}} 
\def\gsim{\mathrel{\rlap{\lower3pt\hbox{\hskip1pt$\sim$}}
     \raise1pt\hbox{$>$}}} 

\def\barc{$\bar c c\,\,\,$}
\def\Jp{$J/\psi$ }

\documentclass[twocolumn,letter,showpacs,preprintnumbers,amsmath,amssymb,floatfix,nofootinbib]{revtex4}

\usepackage{graphicx}
\usepackage{dcolumn}
\usepackage{bm}

\begin{document}


\title{Charmonium in strongly coupled
  Quark-Gluon Plasma}

\author{  Clint Young and Edward Shuryak}
\address { Department of Physics and Astronomy\\
State University of New York, Stony Brook, NY 11794-3800 }

\date{\today}
\begin{abstract}
The growing consensus that a strongly-coupled quark-gluon plasma (sQGP) has 
been observed at the SPS and RHIC experiments suggests a different framework 
for examining heavy quark dynamics. 
We present both semi-analytical 
treatment of Fokker-Planck (FP) evolution in pedagogical examples  
and numerical Langevin simulations of
evolving $c \bar c$-pairs on top of
a hydrodynamically expanding fireball. In this way, we may conclude that the 
survival probability of bound charmonia states is greater than previously 
estimated, as the spatial equilibration of pairs proceeds through a 
``slowly dissolving lump'' stage related to the pair interaction.
\end{abstract}
\maketitle
\begin{narrowtext}

\section{Introduction}
\subsection{Overview}
Charmonium suppression is one of the classical probes used in heavy 
ion collisions. Since charm quark pairs originate during early hard 
processes, they go through all stages of the evolution of the system. A 
small fraction of such pairs $\sim O(10^{-2})$ 
produces bound $\bar c c$ states. By comparing
the yield of these states in AA collisions to that in pp collisions 
(where matter is absent) one can observe their survival probability, giving
us important information about the properties of the medium.

Many mechanisms of $J/\psi$  suppression
in matter were proposed over the years.
The first, suggested by one of us in 1978 \cite{Shu_QGP}, is
(i) a  gluonic analog to ``photo-effect''
 $g J/\psi\rightarrow \bar c c$. Perturbative calculations
of its rate  
\cite{Kharzeev_Satz} predict a rather large 
excitation rate of the charmonium ground state. Since charmonia
are surrounded by many gluons in QGP, this
leads to a conclusion by  Kharzeev
and Satz\cite{Kharzeev_Satz} that nearly all charmonium
states at RHIC should be rapidly destroyed.  
If so, the observed   $J/\psi$ would come mostly from
recombined charm quarks at chemical freezout, as advocated
in \cite{Andronic:2003zv}.

However this argument is only valid in weakly-coupled QGP,
in which
the charm quarks would fly away from each other as soon as enough
 energy 
is available. As we will show below,
in $strongly$-coupled QGP (sQGP)  propagation of charmed quark
is in fact 
very different. Multiple momentum exchanges with matter will lead to
rapid equilibration in momentum space, while motion in position space 
is slow and diffusive in nature. Persistent attraction between quarks   
makes  the possibility of returning back to the ground state for  the
$J/\psi$ quite substantial, leading to a substantial survival 
probablity even after several fm/c time in sQGP.
 
Another idea (ii) proposed by 
Matsui and Satz  \cite{MS}  
focuses on the question of whether
charmonium states do or do not survive {\em as a bound state}. 
They argued that  
because of the deconfinement and the
Debye screening, the effective \barc  attraction in QGP 
is simply too small to hold them together as
bound states. Quantum-mechanical calculations by Karsch et al \cite{KS}
and others have used the free energy, obtained from the
lattice, as an effective potential (at $T>T_c$)
\be F(T,r)\approx -{4 \alpha(s)\over 3 r }\exp(-M_D(T)r)+F(T,\infty)
\label{eqn_F} \ee
They have argued that as the Debye screening radius $M_D^{-1}$
decreases with $T$ and becomes smaller than the root mean square radii
of corresponding states $\chi,\psi',J/\psi,\Upsilon'',\Upsilon',\Upsilon...$, 
those states should subsequently melt. Furthermore, it was found that
for \Jp the melting point is nearly exactly $T_c$, making it
a famous ``QGP signal''.

These arguments are correct asymptotically at high enough 
$T$, but the central issue  is what happens  at (so far 
experimentally accessible at RHIC) $T=(1-2)T_c$.
Dedicated lattice studies \cite{quarkonia} extracted quarkonia spectral
densities using the so called maximal entropy method (MEM) to 
analyze the temporal (Euclidean time) correlators.
Contrary to the above-mentioned predictions, the peaks
corresponding to $\eta_c,J/\psi$ states remains basically
unchanged with $T$ in this region,  indicating the
dissolution temperature is as high as $T_{\psi}\approx (2.5-3)T_c$.
Mocsy et al  \cite{Mocsy:2006qz} have used the Schr\"odinger equation for 
the Green function in order to find an effective potential which would 
describe best
not only the lowest s-wave states, but the whole spectral density.
Recently \cite{Mocsy:2007bk} they have argued that a near-threshold
enhancement is hard to distinguish from a true bound state: 
according to these authors, the  above mentioned
MEM dissolution temperature is perhaps too high.

Another approach to charmonium in heavy-ion collisions, taken by Grandchamp 
and Rapp \cite{Grandchamp:2002iy}, does not rely on the perturbative 
calculation of the excitation cross-section.
Charm rescattering is enhanced by formation of bound states
in QGP.
 \Jp lifetimes were calculated at various temperatures using 
heavy quark effective theory \cite{vanHees:2004gq}:
the resulting widths are typically a few hundred MeV.
If so,  the total number of rescatterings of \Jp in the fireball
during its lifetime ($\sim 10 \, \fm/c$) is large (10-30).  
This model still has fairly large cross-sections for \Jp-annihilation, 
so in their so-called two-component model, many of the final charmonia measured 
are required to originate from statistical coalescence of single charm in the 
plasma into charmonium states.

There are also other quantum-mechanical studies of the issue
since the pioneering paper \cite{KS}.
Zahed and Shuryak \cite{SZ_bound} argued 
that one should $not$ use the free energy $F(T,r)$ 
as the effective potential, because it corresponds to
a static situation in which infinite time is available
for a ``heat exchange''  with the medium.
In  the dynamical real-time situation they proposed to think in terms
of level crossing and  Landau-Zener formalism, widely used in various
quantum-mechanical applications.
In the ``fast'' limit (opposite to the ``adiabatically slow'' one)
all  level crossings should instead  be ignored.
This corresponds to retaining
pure states (described by a wave function rather than density
matrix) without
the {\em  entropy term} in $F$, which is nothing else but 
the internal energy  
\be V(T,r)=F(T,r)-TdF(T,r)/dT=F+TS \ee instead,
as an alternative effective potential.
Such potential $V(T,r)$ 
(extracted from the same lattice data)
leads to much more stable
bound states, putting charmonium melting temperature to higher $T\sim 3T_c$.
A number of authors \cite{Wong_Alberico} have used  effective
potentials in between those two limiting cases. 
However, as it will be clear from what follows,
we think it is not the bound states themselves which
are important, but kinetics of transitions between them.
In a nutshell, the main issue is {\em how small is the separation in the 
\barc pair when the QGP is over}, not in which particular states
they have been during this time.

The heavy $Q \bar Q$ potential depends not only on the temperature but also
on the velocity of the \barc pair relative to the medium.
This effect has been studied e.g. by means of the AdS/CFT
correspondence in \cite{Ejaz:2007hg} and it was found that
the bound state should not exist above a certain critical velocity.
So, if the existence of a bound state is truly a prerequisite for
$J/\psi$ survival, one would expect additional suppression
at large $p_t$. This goes contrary to the well-known formation-time argument
\cite{Karsch:1987uk} and the experimental
evidence, indicating the disappearance of the  $J/\psi$
suppression at large $p_t$.
We think this as a good example of how
important the real-time dynamics of the \barc pair in medium is:
and indeed we are going to follow it below from its birth to
 its ultimate fate. 

Let us now briefly review the experimental situation. 
For a long time it was dominated by the  SPS experiments
NA38/50/60, who have observed both ``normal'' nuclear absorption
and an ``anomalous'' suppression, maximal in central  Pb+Pb
collisions \cite{NA60}. Since at RHIC QGP has a longer lifetime
and reaches a higher energy density, 
straightforward extrapolations of the naive \Jp melting scenarios 
predicted near-total suppression. And yet, the first RHIC data
apparently indicate a survival probability similar to that 
observed at the SPS.

One possible explanation \cite{J/psi_recombination} 
is that the \Jp suppression 
is cancelled by a recombination 
process from unrelated
(or non-diagonal)  $\bar c c$ pairs floating in the medium.
However this scenario needs quite
accurate fine-tuning of two mechanisms.
It also would require rapidity and momentum distributions  of 
the $J/\psi$  at RHIC to be
completely different from those in a single hard production.

 Another logical possibility \cite{Karsch:2005nk} 
is that the $J/\psi$ actually does survive  both 
at SPS and RHIC: while  the anomalous 
suppression observed may simply be due to suppression 
of higher charmonium states, $\psi^{'}$ and
 $\chi$,  feeding down about 40\% of  $J/\psi$
 in pp collisions. These authors however have not
attempted to explain why  $J/\psi$ survival probablity
can be close to one.

This is precisely the goal of 
the present work, in which we  study
 dynamically $how$ survival of $J/\psi$ happens.
We will see that it
 is enhanced by two  famous signatures
of sQGP, namely (i)  a {\em very small
charm diffusion constant} and  (ii) 
{\em strong mutual attraction} between
charmed quarks in the QGP. We found that
 \Jp survival through
the duration of the QGP phase  $\tau\sim 5 \fm/c$
is about a half.

The sequence of events can be schematically described as a four-stage
process
\be 
{{\Longrightarrow} \over {\rm{(\bar c c \, production)}}}
\,\, f_{\rm{initial}}
 \,\,{{\Longrightarrow} \over {\rm{(mom. relaxation)}}}
\,\,f_{\rm{quasi-equilibrium}}\ee
$$ \,\,{{\Longrightarrow} \over {\rm{(leakage)}}}
 \,\,f_{\rm{final}}
 {{\Longrightarrow}
  \over {\rm{(projection)}}}
\,\,  J/\psi $$
A new element here is a  two-time-scale evolution,
 including
rather rapid momentum
 relaxation to a quasiequilibrium distribution 
which differs from the equilibrium one  at large enough distances.

\subsection{Charmonium  potentials at $T>T_c$}
The interaction of the \barc pair will play a significant role in 
this paper, and thus we  briefly review what is known about
them. The details can be found in the original lattice results:
we will point out only the most important qualitative features.

Perturbatively, at high $T$ one expects a  Coulomb-like
force, attractive in the color singlet and repulsive in
the color octet
channel, with the relative strengths 8:(-1) (so that color average will
produce zero effect). As shown by one of us 
many years ago\cite{Shu_QGP},
the Coulomb forces are screened by the
gluon polarization operator at distances $\sim 1/gT$. 

Quantitative knowledge of the interaction
comes from large set of lattice measurements of the
free energies associated with a pair of heavy quarks 
in an equilibrium heat bath. These data include both results
by the Bielefeld-BNL group and in dynamical QCD with $N_f=2$ by 
Aarts et al. \cite{Aarts:2007pk}.

At $T>T_c$ one is in a deconfined phase, so at large 
quark separations one expects effective potentials 
to go to a $finite$  $V_\infty(T)$. Yet when 
the value of this potential significantly
exceeds  the temperature, the actual probablity of quark separation
is small  $\sim \exp(-V_\infty(T)/T)$. 
As we already mentioned in the preceding section, 
the appropriate  potential for dynamical \barc pair
is not yet definitely determined, with suggestions ranging
from free energy to potential energy measured on the lattice.  

The difference between the two 
-- the entropy associated with \barc pair -- 
is very large near $T_c$, reaching the value $S\sim 20$
at its peak.
It means that a huge number of
excited states $\sim \exp(S)\sim \exp(20)$ would be excited
in adiabatically slow motion of the pair. We think
that in realistic motion of \barc pair much less
states are actually excited, and thus, following \cite{SZ_bound},
we will use the potential energy instead of free energy.
A cost of this is much larger potential barrier,
 reducing \barc dissociation. Indeed, $V_\infty(T)$
 near $T_c$ is  large, 
reaching about 4 GeV at its peak.

For the simulation we need a parametrization of the heavy quark-antiquark 
potential above deconfinement as a function of temperature and separation. 
We use the same parametrization as in \cite{Mocsy:2006qz}:

\begin{figure}
\label{fig_pot}
\includegraphics[width=8cm]{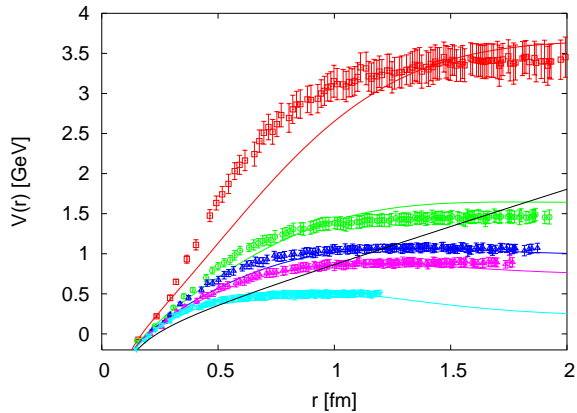}
\caption{ (Color online.)
Parametrization of the potential for lattice data from 
\protect\cite{Kaczmarek:2005zp}. 
From top to bottom: the potential at $T/T_c=1.02, 
1.07, 1.13, 1.18, 1.64$ . }
\end{figure}

\be V(r,T) = -\alpha \frac{e^{-\mu(T)r^2}}{r}+\sigma r
e^{-\mu(T)r^2}\\ \nonumber
+C(T)(1-e^{-\mu(T)r^2}) \ee
and fit to quenched-QCD lattice data in a temperature range of $1.02T_c$
to $1.64T_c$, assuming $\sigma$ is constant in temperature, $\mu$ varies 
linearly with temperature over this range, and 
$C(T)\propto (T/T_c-0.98)^{-1}$, so that it peaks sharply at $T_c$ as 
$U_{\inf}$ does. The result shown in Fig.\ref{fig_pot}(a) is for 
\be \mu(T) = (0.03+0.006T/T_c)\;\GeV^2 \ee
\be \sigma = 0.22\;\GeV^2 \ee
\be C(T) = \frac{0.15\GeV}{T/T_c-0.98} \ee
with $\alpha$ set to $\frac{\pi}{12}$.

As you can see from Fig.\ref{fig_pot}(a), this fit is not 
perfect. While it is easy to fit a 
function to a single temperature's data set, it is hard to find an adequate 
fit across temperatures. This fit however will prove sufficient, especially 
since it is relatively good for the small separations of  interest.

The classical Boltzmann factor $\exp(U(r)/T)$ for a Coulomb potential
leads to a non-normalizable distribution. Quantum mechanics prevents this, 
which can be crudely modeled by
an effective potential
\be V_{eff}= \hbar^2/Mr^2+V(T,r)\ee
which includes the so called localization energy. 
Dusling and one of us \cite{Dusling:2007cn} have determined more
accurate effective quantum potentials, following
ideas of Kelbg and others and performing path
integrals. Perhaps those should be used in future more
sophisticated simulations. 
In the present simulations we simply turn off the force 
 below $0.2\;\fm$, approximating the effective potential
by a constant.

During the simulation, we allow our 
pairs to exist as either color singlets or color octets. While in a 
zero-temperature pp collision, confinement completely suppresses 
a \barc pair's probability of existing in a color octet state, in the 
deconfined phase this possibility must be considered. We initially create 
a 1:8 ratio of color singlet to color octet pairs, as is expected 
statistically, and then during each timestep we decide if the pair 
exists as a color octet by comparing a random number between zero and one to 
$\frac{1}{Z}\exp(-(U_8-U_1)/T)$, with the color singlet and octet potential 
energies determined from \cite{bielefeld}. In other words, the pairs exist in 
thermal equilibrium in color space at all times. When a pair exists as a color 
octet, they do not interact in our simulation as the spatial variation of the 
octet free energies is quite small. 

Naively one would expect this to create a large difference with results where 
color singlet is assumed for all pairs, however this is not the case. One sees 
this by noticing that for the temperatures and distances of our interest 
(distances of separation for pairs likely to go into the $J/\psi$-state), 
$\frac{U_8-U_1}{T} \sim 10$, meaning that the octet state is suppressed by 
orders of magnitude.

Lattice potentials do not follow the
 simple perturbative relation $V_1=-8V_8$ between singlet 
and octet potentials mentioned above.
While the color singlet channel displays a significant attraction,
the color octet channel has a potential 
which is remarkably flat ($r$-independent). 
On the basis of this feature of the lattice
data we will ignore the 
force in the octet channel in the simulation.  

\subsection{Charm diffusion constant}

Loosely speaking, the effect we are after in this work can
be express as follows: the medium
 is trying to prevent the outgoing
quarks from being separated. 
 It has been conjectured by one of us that
charm quark pairs should get stopped in QGP \cite{Shuryak_stuck_c}. 
RHIC single electron data
suggest that charm quarks, and probably b quarks as well,
are indeed equilibrating much more effectively than
it was thought before. 

The first question one should address is
how a charm-anticharm pair moves in sQGP, and what is the probablity 
to find the \barc in close proximity after time $t$.
Charm quarks are subject of three forces:
(i) the $drag$ force, trying to reduce the difference
between quark momentum distribution and that in
(local) equilibrium quark and matter velocities \\
(ii)  $stochastic$ (or Langevin) force from a heat bath, leading to 
thermal equilibration in momenta and spatial diffusion of the charm quarks \\
(iii) the \barc mutual interaction.
For pedagogical reasons
it is useful to include them subsequently,
first for a stationary non-floating matter at  fixed temperature
(using the Fokker-Planck equation) and then for
a realistic nuclear geometry with a hydrodynamically expanding fireball.

But before we do so, let us briefly remind why the case of heavy
(charmed) quarks is so special. (More detailed discussion of that can be
found in Moore and Teaney \cite{MT}.)
A collision of a heavy quark
with  quasiparticles of QGP leads to change in its momentum
$\Delta p_{HQ} \sim T$, so that the velocity is changed
by a small amount
\be \Delta v_{HQ} \sim T/M_{HQ} \ll 1 \label{vhq} \ee 
Therefore the velocity of the charm quark 
can only change significantly as a result of multiple 
collisions, in small steps. Thus the process can be described via 
appropriate differential equations
such as the Fokker-Planck or the Langevin equations.
Similarly one can argue that 
spatial diffusion of a heavy particle can also be described
in this way,  
because the change in position between 
collisions  are small and uncorrelated.

An assumption necessary for Langevin dynamics to hold is that
the ``kicks'' are random (uncorrelated). As explained
above, for a single heavy quark it follows from the inequality  $M>>T$,
which guarantees that the quark relaxation time is long compared to the
correlation time in matter. For a \barc pair we need an additional 
requirement, that random forces on $each$ quark can  be
treated as $mutually$ uncorrelated. In order to see how good this
approximation is, one should compare the spatial (equal time) correlation 
length $\xi(T)$ in the medium to the typical distance between quarks
for paths which eventually (at the end of the plasma era) will
become charmonia. 
We will provide two different estimates for the former
quantity, which view QGP either as 
a perturbative ``gas''  or a strongly coupled ``liquid''.

In a perturbative gas of gluons, the mass is small
and in the lowest order momentum distribution is thermal.
Thus the maximum of the momentum distribution is at $p\approx 2.7T$,
about 1 GeV at the initial RHIC condition. The
corresponding $half$ wavelength (the region where the field keeps at least 
its sign)
\be \lambda/2=\pi/p\approx .6 \, \fm\sim \xi(T=.4\, \GeV) \ee   

In the liquid regime quasiparticles do not successfully model the 
degrees of freedom. However we do have phenomenological information about 
spatial correlations from hydrodynamics, which propose the so called 
``sound absorption length'' as a measure above which different matter 
``cells'' decorrelate. It is
\be \Gamma_s={4\eta\over3 s T}\ee
with $\eta/s$, the dimensionless ratio of viscosity to entropy
density. Empirically, RHIC data are well described
if it is of the order of $1/3\pi$ (the AdS/CFT strong-coupling
limit), which suggests a spatial correlation
length one order of magnitude smaller $\xi(T=.4\, \GeV)\sim 0.05\, \fm$.

Since the distance between $\bar c$ and $c$ which eventually become
$J/\psi$ is about .5 fm, and since there are good reasons
to believe the latter ``liquid'' estimate of $\xi$ is closer
to reality, we conclude that the main 
Langevin assumption -- the independence
of random forces for  $\bar c$ and $c$ -- is well justified.
Furthermore, one may  think that the same assumption
would even hold for $\bar b$ and $b$, although with worse accuracy.

Since this small parameter in Eq.\ref{vhq} is central to what follows, let 
us remind for orientation of the reader that
at RHIC we speak about the ratio for charmed quark $T/M_c=(1/6-1/5)$,
or $T/M_b=(1/20-1/15)$ for $b$ quark.

Although 
RHIC experiments with charm quarks include direct reconstruction
of charmed mesons $D,D^*$ by the STAR collaboration,
so far the existing vertex detectors are not sufficient
for doing it effectively (upgrades are coming).
Therefore the most
relevant data on charm is based on
observation of  single electrons from heavy quark weak semileptonic
decays. Apart of electromagnetic backgrounds, we do not really
know whether electrons come from $c$ or $b$ decays: it is
believed (but not yet proven) that the boundary between two regimes
is at $p_t\approx 4\, \GeV$. 
Two experimentally observable quantities  are (i) the 
charm suppression relative to the parton model (no matter)  $R^e_{AA}$ 
and (ii) the
azimuthal asymmetry of the electrons relative to impact parameter
$v^e_2=<cos(2\phi)>$. 

Several theoretical groups have analyzed these data, in particular
Moore and Teaney \cite{MT} provided information about
the diffusion constant of a charm quark $D_c$ by 
Langevin simulations. A conclusion following from this work
is that both  $R^e_{AA}$ and $v^e_2$ observed
at RHIC can be described by one value for the charm diffusion
constant in the range 
\be \label{eqn_Dc} D_c\;(2\pi T)=1.5-3. \ee
This can be compared with the 
perturbative (collisional) result at small $\alpha_s$
\be  D^{pQCD}_c\;(2\pi T)=1.5/\alpha_s^{2}. \ee
Assuming that the perturbative domain starts\footnote{Recall
that at $4/3\alpha_s=1/2$ two scalar quarks should fall towards each other,
according to the Klein-Gordon equation: so this is clearly not a 
perturbative region.} 
somewhere at $\alpha_s< 1/3$ one concludes that the empirical value
(\ref{eqn_Dc}) is an order of magnitude smaller than the perturbative 
value.

There are two studies of the diffusion constant at strong coupling. 
One comes from AdS/CFT correspondence~\cite{Maldacena:1997re} (and
thus the results are for
 $\cal N$=4 supersymmetric Yang-Mills theory): the final
expression found by Casalderrey-Solana and
Teaney~\cite{Casalderrey-Solana:2006rq} is
 \be D_{HQ}={2 \over \pi T \sqrt{g^2 N_c}}  \ee
It is nicely consistent (via the Einstein relation) 
with the calculated  drag force
\be \label{eqn_drag_adscft} 
{dP\over dt}= -{\pi T^2\sqrt{g^2 N_c} v \over  2\sqrt{1-v^2}} \ee
by Herzog et al \cite{Herzog:2006gh}. 
One assumption of this calculation 
is that the 't Hooft coupling is large $g^2 N_c\gg 1$,
which means the diffusion constant is parametrically small,
much less than  momentum diffusion in the same theory
$D_p=\eta/(\epsilon+p)\sim 1/4\pi T$.
This result is also valid only for quarks heavy enough 
\be M_{HQ} > M_{eff}\sim
\sqrt{g^2 N_c} T \ee
which only marginally holds for charm quarks.

Let us see what these numbers mean for RHIC (assuming that they are
valid for QCD). The 
'tHooft coupling $g^2 N_c=\alpha_s 4\pi N_c\approx 20-40$
is indeed large, while $M_{\rm{eff}} \approx 1-2 \, \GeV$ is not really
small as compared to the charm quark mass: thus the derivation
is only marginally true. Yet we proceed and get 
\be D_c\;(2\pi T)=4/\sqrt{g^2 N_c}\sim 0.5-1, \label{eqn_adscft_Dc}
\ee  which is in right ballpark of phenomenological numbers.

Another approach to transport properties of a strongly coupled plasmas
is classical molecular dynamics. A classical non-Abelian
model for sQGP was suggested by Gelman et
al \cite{GSZ}, and  recently Liao and Shuryak \cite{LS_monop}
have added magnetically charged quasiparticles.
Those calculations also find that
 the diffusion constant $D$ rapidly decreases as a function
of the dimensionless coupling constant
$\Gamma$ as  a power 
\be D\sim \left({1 \over \Gamma}\right)^{0.6-0.8} \ee
in a liquid domain $\Gamma=1-100$. Qualitatively
it is similar to the $1/\sqrt{g^2N_c}$ of the AdS/CFT result (\ref{eqn_adscft_Dc}).
Unfortunately, at this moment there is no deep understanding
of the underlying mechanism of both strong coupling calculations.

But this is well beyond the aims of the present work:
in what follows we will use $D_c\;(2\pi T)=1.5, 3$ 
as a range for our best current guess.

\section{Charmonium in sQGP: Fokker-Planck formalism}

\label{charmthermal}

Before we described realistic
Langevin simulations of RHIC collisions
it is useful to see first the basic features analytically. 
In this section we describe how (1) the high drag 
coefficient $\eta_c$ causes rapid thermalization of the initially hard 
momentum distribution from PYTHIA, and (2) the small diffusion coefficient
$D_c$, inversely proportional to $\eta_c$, combined with (3) the use of 
the static $Q\bar Q$ internal energy instead of the free energy for the 
pair interaction leads to a slow dissolution of a large peak in the 
distribution in position space, causing less suppression than expected by 
more naive models.
 
To see the first feature, rapid thermalization in momentum space, consider 
a \barc pair with the quarks emerging back-to-back from a hard process with 
momentum $\mmntm>>T$, therefore having relative momentum 
$\Delta\mmntm=2\mmntm$. For $\mmntm>>T$, as is the case initially for 
charm quarks created at the RHIC, we may neglect the random kicks that 
the quarks experience from the medium and consider only the drag:
$\frac{d(\Delta\mmntm)}{d\tau}=-\eta_c\mmntm$, where 
$\eta_c=\frac{T}{M_cD_c}$. This leads to a very simple formula for the 
relative momentum of a pair at early times:
\be \Delta \mmntm(t)=\exp(-\eta_c \tau ) \Delta \mmntm(0)\;. 
\label{eqn_drag_only} \ee
Using the AdS/CFT value for the diffusion coefficient, this leads to a 
drag coefficient $\eta_c \approx .6\;\GeV$.

Let us now examine the distribution of \barc pairs created 
with PYTHIA pp-event generation at energies of 200 GeV (see more on this
in Section \ref{sec_init}). The initial transverse momentum distribution 
\footnote{The initial $rapidity$ distribution is wide also, but since 
longitudinally Bjorken-like hydrodynamics starts immediately, a charm quark 
finds itself with comoving matter with the same rapidity, and thus has little 
longitudinal drag. Transverse flow is slow to be developed, thus there is 
transverse drag.} is broad compared to the thermal distribution, and 
therefore we may apply Eq. \ref{eqn_drag_only} for early times. We 
parametrize the initial relative momentum distribution as in Section 
\ref{sec_pyth} and replace $\Delta \mmntm$ with $\Delta \mmntm \exp(\eta_c
\tau )$, which is the formula for a pair's $initial$ relative momentum in 
terms of its relative momentum at proper time $\tau$. Next, we compute the 
overlap of this distribution at early times with the \Jp-state's Wigner 
quasi-probability distribution to determine the probability of a random 
\barc taken from this ensemble to be measured as a \Jp particle, an 
approach detailed in Appendix \ref{prob}.

Fig. \ref{pyth_proj_Jpsi} shows this probability as a function of time for 
$\eta_D=0.88$. The initial value of the projection is 0.8\%, on the same 
order of magnitude as the experimental value of 1\% for 
$\frac{\sigma_{J/\psi}}{\sigma_{c\bar c}}$ obtained from \cite{sjp},
\cite{scc}. The projection increases
as the PYTHIA-with-drag distribution narrows but then drops when the width 
shrinks more than the $J/\psi$'s width in momentum space and as the quark 
pair's separation increases. Once the time reaches about 1 fm/c, the 
probability for a pair to go into a $J/\psi$-state is about where we 
started, and we quit looking at this approach after this much time because 
the mean transverse momentum for a quark is the thermal average.

\begin{figure}
\label{pyth_proj_Jpsi}
\includegraphics[width=8cm]{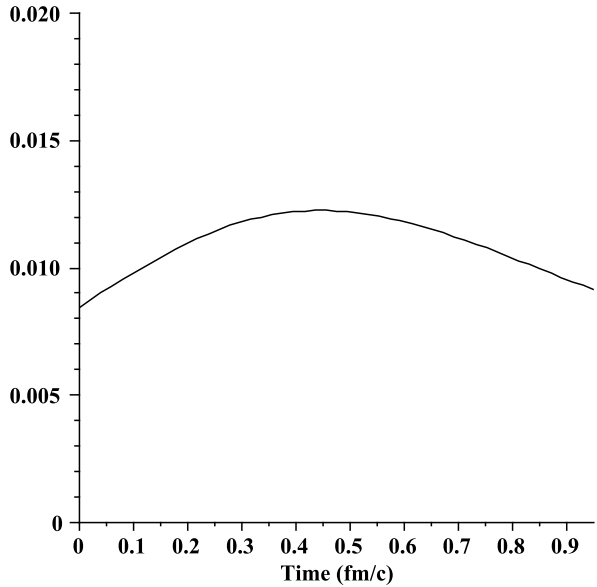}
\caption{Probability of a \barc pair going into the $J/\psi$-state vs. time, 
for very early time.}
\end{figure}

After this first 1 fm/c of the QGP phase, the \barc distribution has 
thermalized in momentum space and the evolution in position space (diffusion) 
needs to be examined. The root mean square distance for diffusive motion is 
given by the standard expression 
\begin{equation} 
\llangle x^2 \rrangle=6 D_c \tau 
\end{equation}
where $\tau$ is the proper time and the interaction between the quarks has 
been neglected. The ``correlation volume'' in which one finds
a quark after time $\tau$ is
\be V_{\rm{corr}}={4\pi \over 3}(6 D_c \tau)^{3/2}\ee
and one may estimate for the probability of the \barc pair to be measured 
in the \Jp-state as
\be \label{eqn_prob_survival}
P(\tau)\sim R_{J/\psi}^3/(6 D_c \tau)^{3/2} \ee
So neglecting the pair's interaction leads to a small probability that 
\Jp-states will survive by the hadronization 
time at the RHIC ($\tau\sim 10\,\fm/c$), even for small values of the 
diffusion coefficient.

To get an idea for how this simple result is changed by the inclusion 
of an interaction between the constituent quarks in a given \barc-pair, let 
us examine the Fokker-Planck equation for the \barc distribution in 
relative position:
\be \label{eqn_FP}
{\partial P \over \partial t}=D {\partial \over \partial\pstn}
 f_0 {\partial \over \partial\pstn} (P/f_0)  \ee
where $f_0(r)\propto exp(-V_{\bf eff}(r)/T)$ is the $equilibrium$ 
distribution
in the magnitude of relative position $r$. 
By substituting the potential shown above at $T=1.25Tc$
and  $D_c\times(2\pi T)=1$
into the Fokker-Planck equation (for demonstration
in a single spatial dimension only) we solve it numerically and
find how the relaxation process proceeds.
A sample of such calculations is shown in Fig. \ref{fp_numeric}.
It displays two important features of the relaxation process:\\
(i) during a quite short time $T, 1\, \fm$ the
initial distribution (peaked at zero distance)  relaxes locally to
the near-equilibrium distribution 
with two peaks, corresponding to optimal distances of the
equilibrium distribution $f_0$,  where the effective
potential is most attractive; \\
(ii) the second stage displays a slow ``leakage'', during which
the maximum is decreasing while the tail of the distribution
at large distances grows. It is slow because the right-hand side
of the Fokker-Planck equation is close to zero, as the distribution
is nearly $f_0$. The interaction drastically changes the evolution of the 
\barc distribution in position space, and this will be demonstrated again 
in the full numerical simulation of the next section.

\begin{figure}
\label{fp_numeric}
\includegraphics[width=8cm]{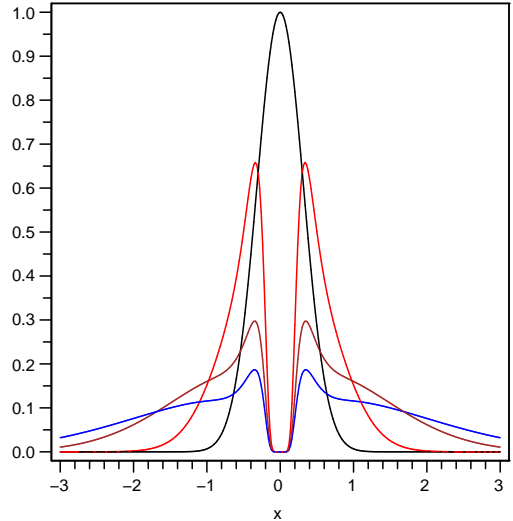}
\caption{ (Color online.) Numerical solution of the one-dimensional 
Fokker-Planck equation for an interacting \barc pair. The relaxation
of the initial narrow Gaussian distribution
is shown by curves (black, red,brown,green,blue, or top to bottom
at r=0) corresponding to times $t=0,1,5,10\;\fm$, respectively.}
\end{figure}

\section{ Langevin evolution of the $\bar c c$ pairs }

\label{sec_init}
\subsection{ Langevin evolution in static medium: quasiequilibrium}
Before we turn to expanding fireball, we first study
the evolution of  $\bar c c$ pairs at fixed $T=1.5T_c$ and in the absence
of hydrodynamical flows. The first thing we would like to demonstrate is 
the strong influence of the heavy quark interaction. 
The resulting distribution over interquark separation at time 
$10\;\fm/c$
 are shown in Fig. \ref{t=9}, with the interaction (red squares) and without
 (blue triangles).
 The value for a given pair's separation $r$ is 
weighted by $\frac{1}{r^2}$ so that the spatial phase space of the 
distribution is divided out.
With the interaction on, we find the same behavior of a ``slowly dissolving 
lump'' as that seen in the solution of the Fokker-Planck equation
when the interaction is ``turned on''.

\begin{figure}
\label{t=9}
\includegraphics[width=8cm]{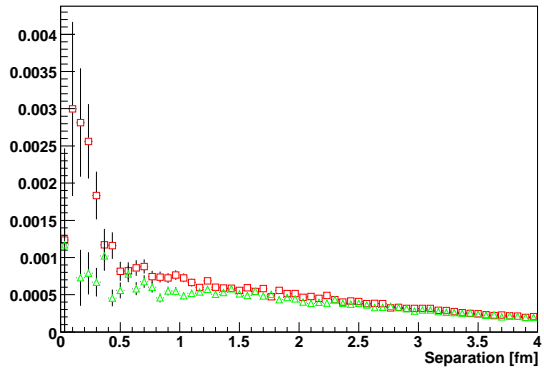}
\caption{(Color online.) Distribution over quark pair separation at 
fixed $T=1.5T_c$ after 9 fm/c, with (red squares) and without (green triangles)
the \barc potential.
}
\end{figure}

Further study of this has shown convergence of 
its shape to a particular one, which 
persist for a long time and which we would call
``quasiequilibrium''\footnote{This situation should not be confused
  with stationary nonequilibrium solutions of the Fokker-Planck
equation, in which there is a constant flow through the
distribution because of matching source and sink.}.
While the true equilibrium of course corresponds to complete
dissolution of a single \barc pair, it turns out that leakage to large 
distances
affects the distributions of separation and energy in normalization only.
In Fig. \ref{etot_pet_fit} is the energy 
distribution for the ensemble of pairs at $\tau=9\;\fm/c$ after evolving 
under Langevin dynamics at a fixed temperature $1.05T_c$, and it is shown to 
be the same distribution, up to normalization and statistical uncertainty, 
as the distribution reached by the pairs in a full heavy-ion simulation of 
the most central collisions.

\begin{figure}
\label{etot_pet_fit}
\includegraphics[width=8cm]{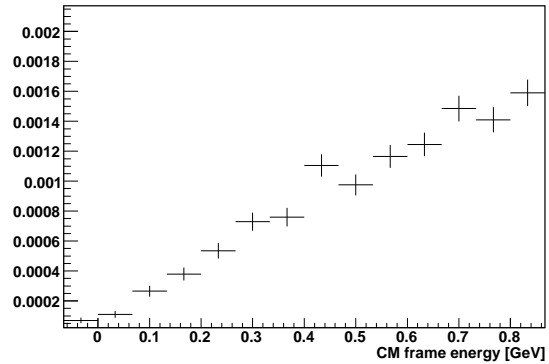}
\caption{
Energy distribution of the $c \bar c$ pairs 
(in the center of mass frame of the pair)  after
Langevin evolution at a fixed temperature $T=1.05T_c$ for a time $t=9\;\fm/c$
long enough for quasiequilibrium to be reached.
}
\end{figure}

We show the energy distribution in this region because
it is related to a very important issue of the charmonium
production, namely production of $\psi^{'},\chi$ states and 
subsequent feeddown into the $J/\psi$. When a quasiequilibrium
distribution is reached, the
production {\em ratios}  of charmonium states
are stabilized at thermal (statistical) model values, in
spite of the fact that the
overall normalization continues to shrink due to leakage into infinitely large
phase space at large distances.

(The energy distribution itself contains a Boltzmann factor
but also the density of states. A model case of purely Coulomb
interaction allows one to calculate it in a simple way: as shown
in Appendix \ref{quantum} we found that in this case the absolute shape
of the quasiequilibrium distribution is reproduced
as well.)

The existence of quasiequilibrium
 is in good correspondence to observations. It was noticed
already a decade ago \cite{Sorge:1997bg}
 that the SPS data on centrality dependence of $N_{\psi^{'}}/N_{J/\psi}$ ratio
approached the equilibrium value (for chemical freezout)
\be {N_{\psi^{'}}\over N_{J/\psi}}=exp(-\Delta M /T_{ch}) \label{eqn_chem}\ee
with the chemical
freezout at $T_{ch}=170\,\MeV$, 
as is observed for ratios of other hadronic species.

One possible explanation of it can be charmonium 
$recombination$ (from independent
charm quarks) at chemical freezout, advocated by
 \cite{Andronic:2003zv} and others.
However our findings show that the same ratio
naturally appears  even for a $single$ \barc
pair dissolving in a thermal medium, in a ``quasiequilibrium''
occurring at the leakage stage. Especially at SPS, when statistical
recombination requires a charm density which is too large, this
is an attractive possibility.

\subsection{Production of the initial $\bar c c$ pairs}

\label{sec_pyth}

 We start with
\barc events produced with PYTHIA, a particle physics event generator 
\cite{sjostrand}. PYTHIA  yields \barc pairs through 
a set of perturbative QCD events: through Monte-Carlo it will select
initial partons from the parton distribution function of one's choosing and 
proceed through leading-order 
QCD differential cross sections to a random 
final state. The $p_t$ and rapidity distributions
of charm produced in pp collisions is believed to be rather adequately
represented.

By using PYTHIA we do not however imply that it solved many open issues
related with charm production, such as color octet versus singlet
state. It also leaves open the very important issue of
$feeddown$ from charmonium excited states (see below).
One more open question -- needed to start our simulations --
is how to sample the distribution in 
position space. Indeed,
each pQCD event generated in PYTHIA is a momentum eigenstate without 
any width, so by Heisenberg's uncertainty relation they are spatially
delocalized, which is unrealistic.

 We proceed assuming the form for the 
initial phase-space distribution to be 
\be P_{\rm{init}}(\pstn, \mmntm) \propto P_{\rm{PYTH}}(\mmntm)
\exp(-\pstn^2/2\sigma^2) \ee
By setting $\sigma=0.3\;\fm$ one can tune energy distribution to give
a reasonable probability for the formation of $J/\psi$ state in 
pp events. 

However this does $not$ yield correct formation
probabilities of $\chi,\psi^{'}$. It is hardly surprising, since
for example  the $\psi^{'}$ wavefunction has a sign change
at certain distances, and the exact production
amplitude is required  for a projection, an
 order-of-magnitude size estimate is not
good enough. Since feeddown from those states contributes about 40\%
of the observed $J/\psi$, we simply refrain from
any quantitative statements about pp (and very peripheral
AA) collisions, focusing only on distributions $after$ 
some amount of time spent in QGP.

\begin{figure}
\label{etot_vs_time}
\includegraphics[width=8cm]{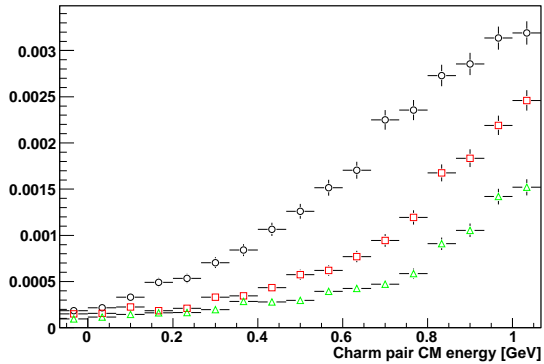}
\caption{(Color online.)
Evolving energy distribution for an ensemble of \barc pairs 
 at time moments $t=2,3,10\,$ fm/c (circles,squares and triangles, respectively).}
\end{figure}

\subsection{Langevin motion of \barc pair in an expanding fireball}

\label{fireball}

Finally, we model the motion of a charm quark pair in an evolving fireball.
We use the same framework and programs
used in \cite{MT} to examine motion of a single charm quark for
propagation of an interacting pair.

We start with large number of \barc produced with PYTHIA pQCD event 
generation, and randomly place them in position space 
throughout the fireball, using  a Monte-Carlo Glauber calculation.

Then, the pairs are evolved in time according to the Langevin equations: 
\be {d \mmntm \over dt}=-\eta \mmntm + {\bf \xi}-{\bf \nabla} U \ee
\be {d \pstn \over dt}={\mmntm \over m_c} \ee
where ${\bf \xi}$ corresponds to a random force and $\eta$ the drag 
coefficient. The condition that the Langevin equations evolve a distribution 
towards thermal equilibrium gives the  relation $\langle \xi_i(t) 
\xi_j(t') \rangle = 2MT \eta \delta_{ij} \delta(t-t')$.
We proceed here with our diffusion constant set by the results of \cite{MT}
to be $\eta_D=\frac{2\pi T^2}{1.5M}$.
and as discussed earlier with our potential as $V$ instead of $F$. 

Now we examine the evolution of the quark pairs as discussed before, 
examining pairs at mid-rapidity in a boost-invariant, 2-dimensional 
ultra-relativistic gas simulation, the same hydrodynamical simulation used 
in \cite{MT}. We stop the Langevin-with-interaction evolution when
$T<T_c$. The distribution over energy at different moments
of the time is shown in Fig.\ref{etot_vs_time}.

We will discuss our results subsequently, at two different levels of 
sophistication. First we will
show results with only the total number of bound states monitored, and then 
we will show results where the different charmonia states and feeddown 
to \Jp are considered and compare these results with PHENIX data. 

How we determine whether or not to call a \barc pair in our simulation bound, 
and what particular charmonium state the pair exists as if it is bound, is 
discussed in Appendix \ref{prob}. Fig. \ref{jpsioccb_y0}
shows the number of ``bound''  \barc pairs
as a fraction of their total number, monitored throughout the course of the 
simulation as a function of time.
Note that realistic hydrodynamical/Langevin simulation 
 agrees qualitatively with the analytic results of Section 
\ref{charmthermal}. During the first fm/c one finds some boosts
 in the probability for a \barc pair to be bound,
due to rapid thermalization in momentum space.  Later
the probablity falls due to the slow diffusion in position space.
This figure emphasizes our main qualitative finding,
the survival probability of $J/\psi$ on the order of a half.

\begin{figure}[t]
\label{jpsioccb_y0}
\includegraphics[width=8cm]{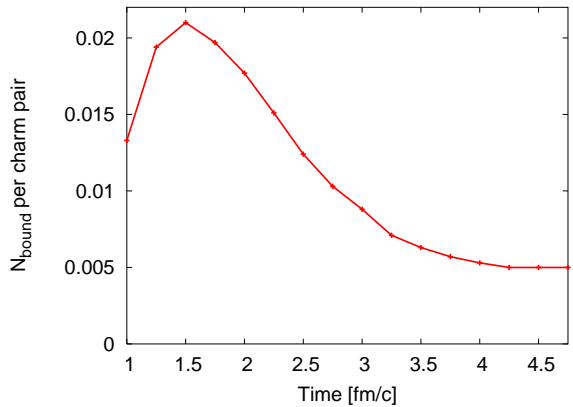}
\caption{ (Color online.)
Probability of \barc pairs to be bound  at 
RHIC Au+Au, $\sqrt{s}=200\;\GeV$, mid-rapidity.}
\end{figure}
 
\subsection{Shadowing and ``normal'' absorption}

Experimental data include
not only the ``QGP suppression'' we study in this work, but also
(i) the initial-state effects
(modified parton distributions in  ``cold nuclear 
matter'') plus (ii) the so called ``normal nuclear absorption''. 
The way we have chosen to display 
 PHENIX data \cite{phenix_nov06} is as follows:
before we compare those with our results we
``factor out'' the cold nuclear matter effects, by 
defining (for any given rapidity $y$) the following 
ratio of Au+Au and d+Au data 
\be R^{anomalous}_{AA}(y)=\frac{R^{PHENIX}_{AA}(y)}{R_{pA}(y)R_{pA}(-y)} \ee
to  be called the ``anomalous suppression''.
In principle those include only data: but 
unfortunately the large dAu sample taken in 2008 is 
not yet analyzed (at the time of this writing), while the 2003 set
has error bars which are too large.
 This forces us to use a model at this point,   
following Kharzeev et al\cite{kharzeev_diss} with
 $R_{pA}=exp(-\sigma_{abs}\langle L \rangle n_0)$, where 
$\langle L \rangle$ is the mean path length of the $J/\psi$ through nuclear 
matter, $n_0$ is the nuclear density, and $\sigma_{\rm{abs}}$ is the nuclear 
absorption cross-section (parametrized from \cite{kharzeev_diss} to be 
$0.1 \fm^2$ for rapidity $y=0$). Finally, for rapidity $y=0$, we rewrite this 
as $(R_{pA}(y=0))^2=\exp(-\sigma_{abs} \llangle N_{part} \rrangle$, 
where $\llangle N_{part} \rrangle$ is the density per unit area of 
participants in the collision plane. We further
used Glauber model calculations  
\cite{kharzeev_glauber} to determine $\llangle N_{part} \rrangle$ for a given 
$N_{\rm{part}}$ at PHENIX. We divide each Au+Au
 data point from PHENIX by this 
quantity and call it $ R^{anomalous}_{AA}(y=0)$ plotted as points
in Figs.\ref{raa_vs_npart} and \ref{plot_feeddown}, to be compared with  
our simulation.

\begin{figure}
\label{raa_vs_npart}
\includegraphics[width=8cm]{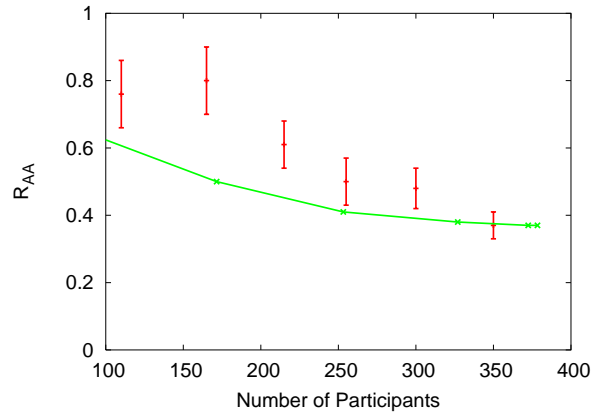}
\caption{(color online)
Points show the magnitude of the anomalous suppression
at mid-rapidity $ R^{anomalous}_{AA}(y=0)$ versus the centrality (the number of
 participants), using Au+Au PHENIX data. The curve
is the probability to be bound (determined by the energy
projection) at the end of the QGP era, when $T=T_c$. }
\end{figure}

\subsection{$\psi^{'}$ production and feeddown}
\label{feeddown}
Next we calculate the 
ratio $N_{\psi^{'}}/N_{J/\psi}$ in our simulations, for different 
centralities. There are well known NA38/50/60 measurements of
this ratio at the SPS, but at RHIC it has been measured so far only 
in pp collisions by the PHENIX detector \cite{:2008as}
 to be 0.14 $\pm$ 0.04, which makes the ratio of direct $\psi^{'}$ to 
$J/\psi$ particles 0.24 as in \cite{Digal:2001ue}. Hopefully higher 
luminosity at RHIC will make possible a future measurement of
 this ratio in Au+Au collisions of 
various centralities.

We calculate $N_{\psi^{'}}/N_{J/\psi}$ as follows: 
(i) first we run our simulation 
and determine the distributions $f(E)$ over the
\barc pair energy $E=E_{CM}-2M_c$
(in the pair center of mass frame); (ii) 
then we compare
those to quasiequilibrium ones from simulations
at  fixed temperature (slightly above $T_c$)  $f_0(E)$.
Since both are done for the same interaction,
in the ratio  $f(E)/f_0(E)$ the density of states
drops out. This ratio tells us how different the actual
distribution is from that in quasiequilibrium.
 (iii) Then we form the 
$double$ ratios, at two
relative  energies corresponding to $\psi^{'}$ and $J/\psi$ masses
(minus $2m_{charm}$)
\be
R_{\psi^{'}/J\psi}=\frac{f(.8\;\GeV)}{f_0(.8\;\GeV)}/\frac{f(.3\;\GeV)}{f_0(.3\;\GeV)} 
\label{double_ratio}
\ee
This now includes nonequilibrium effects for both
of them. Finally (iv) we  switch
from continuum classical distributions to quantum one, assuming
that in quasiequilbrium the relation (\ref{eqn_chem}) holds.
If so, the particle ratio is a combination of nonequilibrium and
equilibrium factors
\be {N_{\psi^{'}} \over  N_{J/\psi}}= R_{\psi^{'}/\psi}\exp(-\Delta
m/T)\ee
The double ratio ( or
 $\exp(\Delta 
M/T)N_{\psi^{'}} / N_{J/\psi}$) is 
plotted vs centrality in Fig. \ref{psiratio}.
As one can see, it goes to unity for the most central
collisions: so  quasiequilibrium
is actually reached in this case. For mid-central bins
the $\psi^{'}$ production is about twice larger because
of insufficient time. This is to be compared
to the experimental pp value for the ratio,
which is about 5. (We remind the reader that
PYTHIA plus our classical projection method does not work
for pp collisions.) \\

\begin{figure}
\includegraphics[width=8cm]{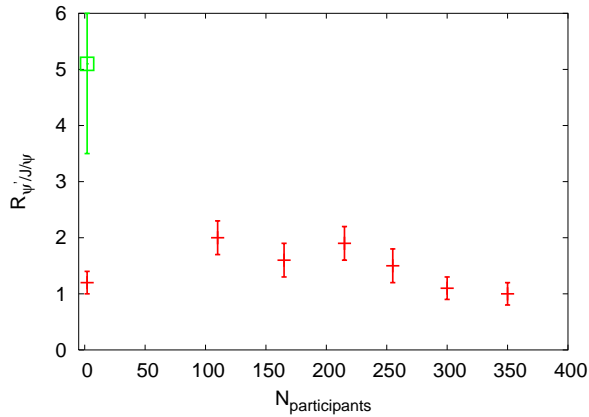}
\caption{(Color online.) The double ratio $R_{\psi^{'}/\psi}$ defined
in (\ref{double_ratio}) versus centrality (number of participants).
One point (green box) at $N_{part}=2$ corresponds to experimental
data for $\psi^{'}$ and the $direct$ \Jp,  for pp collisions.
}
\label{psiratio}
\end{figure}

Finally, we use this result to estimate the effect 
 of feeddown from higher states. 
To do this, we write the final number of $J/\psi$ particles observed 
as the  number of directly produced $J/\psi$ particles plus the number 
of $J/\psi$ particles produced from feeddown from higher charmonium 
states:
\be N^{final}_{J/\psi}=N^{direct}_{J/\psi}\left[1+ R_{\psi^{'}/\psi}
\sum_i (\frac{g_i}{3})
  \exp(-{\Delta 
M_i \over
T})B_i\right] \ee
where $i$ is summed over 
the $\chi_1$, $\chi_2$, and $\psi^{'}$ particles 
which contribute significantly to feeddown, $B_i$ represents their
 branching ratio into $J/\psi+...$, and $g_i$ is the degeneracy of the state 
(divided by 3, the degeneracy of the $J/\psi$),  $\Delta m_i$ is the mass
 difference  between the i-th state and the $J/\psi$. The
  $ R_{\psi^{'}/\psi}$ is the non-equilibrium
factor discussed above: it is factored outside the sum because
it is very similar for all these states.

Now we are ready to discuss centrality dependence of the 
$J/\psi$ production $including$ the feeddown.
We define for each centrality direct $N_{J/\psi}^{direct}(b)$ as 
 the total number of $c \bar c$ pairs in our ensemble with energy 
(in its rest frame) less than $E<2M_{charm}+0.33\;\GeV$.
The feedown gets its dependence on centrality from 
$ R_{\psi'/\psi}(b)$ determined from simulation. 

The absolute normalization of the results deserves special
discussion. We predict the absolute probability
of  $J/\psi$ production, both direct and with feeddown, 
normalized  per \barc $pair$ produced in the same
collisions (e.g. centrality bin). Unfortunately, the total
cross section of charm production is not yet measured 
with sufficient accuracy to normalize results this way.   

The usual way to present these results is 
in the form of the so called $R_{AA}$ ratio, relating production
in Au+Au at given centrality to that in pp, times the theoretical
(Glauber) predictions for the number of hard collisions.
In other words, the parton model for $\bar c c$ is used.
Unfortunately, the experimental data about feeddown in pp are
still uncertain enough to produce sufficiently
 large scale ambiguity. We cannot obtain this
from our theory as well because
 (as explained above) our classical projections do 
not work adequately for pp (and very peripheral)
collisions. Thus we  see at the moment no
sufficiently accurate way of $absolute$ comparison with the data.
Because of that, we 
simply  normalize our results assuming that there is no
QGP suppression 
( meaning $R^{anomalous}_{AA}(N_{part}<100)=1$).
 The results normalized like this 
are  shown in Fig. \ref{plot_feeddown}: we conclude that
although
feeddown is not large, taking it into account 
helps bring the shape of the anomalous suppression
closer to observations.

\begin{figure}
\includegraphics[width=8cm]{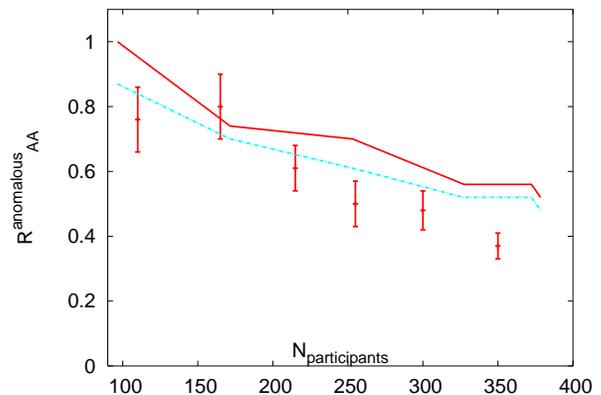}
\caption{(Color online.)
The points are PHENIX data for $ R^{anomalous}_{AA}(y=0)$,
the same as used in Fig.\ref{raa_vs_npart}.
Two curves are our model, with (solid, upper) and without (dashed, lower)
 feeddown.}
\label{plot_feeddown}
\end{figure}

\subsection{Including the ``mixed phase''} 

In our work so far, 
we have only examined the evolution of the $c \bar c$ pairs during the 
QGP phase,  stopping the evolution wherever 
the fluid's 
temperature reached $T_c$. However, in order to understand the 
evolution of charmonia  to their hadronization
 we need to model the 
dynamics of the charm quarks also  through the ``mixed phase'', also
known as the
near-$T_c$ region.

In various hydrodynamic models which describe
heavy ion collisions, the region of roughly $T=0.9-1.1\;T_c$
is treated as a separate ``mixed phase'' distinct from QGP and hadronic
phases.
Indeed, it has a very different equation of state: 
while the temperature and pressure remain
nearly constant, the energy and entropy densities jump by 
a large factor
\footnote{Although the exact nature of matter in the near-$T_c$ region
is not yet understood, let us mention that the original
``mixed phase'' description, originating from the notion of
the first-order phase transition, cannot be true, as ``supercooling''
and bubble formation expected in this case are not observed
experimentally. Lattice gauge theory suggests a smooth crossover-type
transition, with a high but finite peak in the specific heat.
Recently there has been renewed interest in this region, after the so-called
``magnetic scenario'' for it has been
proposed \cite{Liao:2006ry,Chernodub:2006gu}, describing it as a plasma 
containing a significant fraction of magnetic monopoles.}.

What is very important for the present paper
is that the near-$T_c$ region occupies a significant
fraction of the fireball evolution, in spite of being a
very narrow interval in terms of $T$. Indeed,
one should rather focus not on $T$ but on the entropy
density $s$, which shows a simple monotonous
decrease with time $s\sim 1/\tau$ for all three phases.

For a quantitative description of the mixed phase we used
 hydrodynamical calculations, known
to describe radial and elliptic flow well, such as 
the work by Shuryak and Teaney
\cite{Teaney:2001av}. It follows from
their solution that the ``mixed phase''
persists for about $5 \, \fm/c$ after the 
deconfined phase, which is comparable to the
lifetime of the deconfined phase at the very center of the fireball.
Thus it is by no means a small effect, and should
be included in any realistic treatment of a 
heavy-ion collision.

The flow during this time
 was found to be well approximated  
by a Hubble-like expansion with radial velocity $v=Hr$
and time-independent 
$H \approx 0.075 \, \fm^{-1}$ for central collisions.
  For a collision with a nonzero impact parameter
(below we will consider  $b=6.5 \, \fm$), the 
anisotropy of this expansion can be  parameterized
similarly:
\be v_i=H_i x_i\ee
with $i=1,2$ and $H_x=0.078\, \fm^{-1},H_y=0.058 \, \fm^{-1}$: thus anisotropy
is only about 80\% by this late stage. It is 
fair to say that we have a fairly reasonable understanding of how the
 medium flows 
 for these later stages: thus in our simulations we have
used those parameterizations instead of numerical solutions to
hydrodynamics, which were necessary for the QGP phase.

Let us start with two extreme scenarios for the dynamics of 
the charm quarks during this phase of the collision:\\ 1.) the charm 
quarks are completely ``stopped'' in the medium,
 so that they experience the same Hubble-like 
flow as matter; \\ 2.)  $\bar c c$ pairs  do not interact at 
all with the medium near-$T_c$,  moving ballistically
 with constant velocity for the corresponding time 
in the collision. 

If the first scenario were true, the effect of Hubble 
flow would be to increase all momenta of particles
 by the same multiplicative factors 
$p_i(t)=p_i(0)\exp(H_it)$. With sufficiently high drag, Langevin dynamics would
bring the charm quarks rapidly to a thermal distribution, and since $M>>T$ it is a good 
approximation in this case to say that the heavy quarks have been "stopped".
However, we will   show below that at the ``realistic'' 
value used for the drag $\eta_c$ this does $not$ happen
during the time allocated to the mixed phase,
there is instead
 ongoing ``stopping'' of the charm quarks relative to fluid 
elements. (This also will be important for the evolution 
of the azimuthal anisotropy $v_2(p_t)$ for single charm and for charmonium). The second 
scenario predicts $v_2(p_t)$ for single charm quarks which is far smaller than what is measured. 
We do not consider this scenario further even though something might be said for 
modelling charmonium in the mixed phase as interacting far less than single charm.

\begin{figure}[t]
\includegraphics[width=8cm]{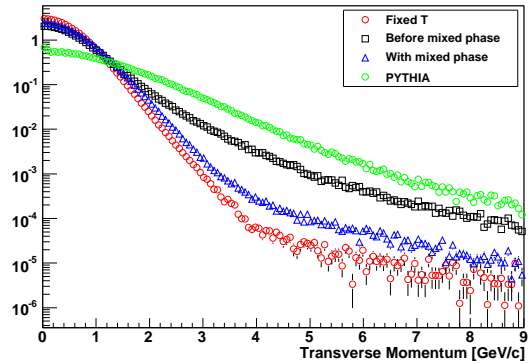}
\caption{(Color online.)
The charm $p_t$-distribution after the mixed phase compared with the 
distribution without flow, the distribution orignating from PYTHIA, 
and the distribution before the mixed phase}
\label{plot_pt_hubble}
\end{figure}

Several  single charm $p_t$-distributions are shown in
Figure \ref{plot_pt_hubble}
(normalized for simplicity to unity).
The initial distribution after hard production  predicted by PYTHIA
is the largest at high $p_t$: this is compared 
to the Langevin evolution before (squares) and
  after (triangles) the mixed phase, for 
a semi-central RHIC collision ($b=7\;\fm$).
In order to see that radial flow still is important, we have also
shown what happens if Langevin evolution happen on top of
unmoving fixed-T plasma (circles). 
This comparison demonstrates once again 
the main point of this paper, that for charm quarks and charmonium
 in a 
heavy-ion collision equilibration is never complete,
even in momentum space:
 so the specific timescales of different phases of matter are 
of fundamental importance.

Unfortunately, in the near-$T_c$ region 
 it is much less clear how to describe the $c-\bar c$
interaction. As we have learned from
lattice data, the difference between free-energy and potential-energy
potentials are very drastic in this case: in the former case
the tension (the force from the linear potential) disappears
while in the latter it becomes about 5 times stronger than
it is in vacuum. As discussed in refs\cite{Liao:2007mj,Liao:2008vj},
the latter is presumably due to a metastable electric flux tube.

Which potential to use depends on timescales of the
$c-\bar c$ dynamics, which are not well understood at this point.
Therefore we took for now a conservative
approach, assuming that  at the near-$T_c$ stage 
charm pairs interact according to the 
simple Coulomb interaction $V=-\alpha_s/r$. Additionally, in our model 
for this phase we assume the interaction of the charm quarks with the 
medium can be modelled with the same Langevin dynamics with the temperature 
approximated as a fixed $T=T_c$ and the flow given as above.
 We found 
that with the simple Coulomb potential used in the mixed phase,
 the survival probability dropped 
slightly but not significantly: and although we do not discuss
other possibilities in this work further, in principle
this can be changed if the potential to be used has significant tension. 

One final interesting observable 
would be a measurement of charmonium  elliptic flow, characterized by
the azimuthal anisotropy parameter $v_2=<cos(2\phi)>$,
induced by ellipticity of the flow on charmed quarks.
A measurement with low statistics has been already made at PHENIX
\cite{Silvestre:2008nk}: both PHENIX and STAR are working on  
 higher statistics data on tape now.
The result of our calculation of 
$v_2(p_t)$, both for single charm
quarks and for the $J/\psi$, is shown in Figure \ref{plot_v2}. 

Greco, Ko, and Rapp also made predictions for the $v_2$ for $D$ and 
$J/\psi$ \cite{Greco:2003vf}, based on a completely different
scenario: in their case  charm 
distributions are completely equilibrated and the charmonium states 
coalesce from them at hadronization.
 In spite of such
difference, our predictions are similar:
$v_2$ of  $J/\psi$ should be less than the $v_2$ of single charm for 
low $p_t$, but then increase past the $v_2$ of single charm 
at $p_t> 3\, GeV$.  This shows, that the observation of charmonium $v_2$
can not be considered the argument for coalescence.

\begin{figure}[t]
\includegraphics[width=8cm]{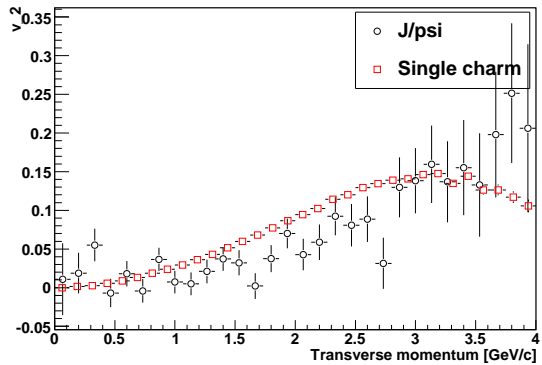}
\caption{(Color online.)
The azimuthal anisotropy versus transverse momentum for both single 
charm and for $J/\psi$.}
\label{plot_v2}
\end{figure}

\section{Summary}

We have studied a relaxation process of a \barc pair
produced by hard processes in heavy ion collision throughout
the sQGP stage using hydrodynamics+Langevin. 
The main elements of the paper are: (i) inclusion of the
interaction force between charm quarks and (ii) emphasis on
deviations from equilibrium
during the finite QGP lifetime. 

Our main finding is that the lifetime of sQGP 
is not sufficient to reach the
equilibrium  distribution of the pairs in space,
allowing for a significant fraction of \Jp $\sim 1/2$ to survive.
This probability 
for charmonium dissociation in sQGP is larger than in
earlier perturbative estimates, or for Langevin
diffusion ignoring mutual interaction. 

That is why there is no large difference between 
suppression at RHIC and SPS, in spite of the
 longer QGP lifetime in the former case.
While the momentum relaxation is rather rapid,
we found that later evolution
reaches the so-called quasiequilibrium regime, which is
  maintained during all time of QGP expansion.
The spatial distributions after some time develop  
a ``core'' in which \barc pairs remain 
in close proximity due to the remaining effective attraction in sQGP, 
combined with a relatively slow leakage into a spreading tail 
toward large $r$. 
We propose quasiequilibrium
 as the clue to an explanation of the apparently thermal ratios
of $\psi^{'}$ and $J/\psi$, especially at SPS. 
 
 The shape of the centrality dependence
of our survival probabality is in agreement with data.
Therefore, although we have not yet directly evaluated
``nondiagonal'' recombination, we  think that
 most of the  \Jp observed at SPS and RHIC
are still from the ``diagonal'' pairs. 
This and other issues will of
course be clarified, as more simulations and data in different
kinematic domains become available.

The calculation is extended to the near-$T_c$ region -- known as a
``mixed phase'' in hydrodynamical calculations. Its duration for
RHIC collisions is about 5 fm/c, comparable to that of the QGP.
We have used simple Hubble-like parameterization and minimal
Coulomb potential, and predict both
 the charmonium $p_t$ spectra and azimuthal
asymmetry parameter $v_2$ which is expected to be measured
soon .

 \vskip 1.0cm

{\bf Acknowledgments.\,\,}
We thank P. Petreczky for providing lattice data on internal energy
$V(T,r)$ used in this work. C. Young would like to thank 
K. Dusling for 
useful discussions on various topics in this work.
We especially thank D. Teaney for permitting 
us use of his hydrodynamics+Langevin 
code and providing much needed assistance. 
This work is partially supported by the US-DOE grants DE-FG02-88ER40388
and DE-FG03-97ER4014.

\appendix

\section{Classical vs. Quantum Dynamics}

\label{quantum}

In this paper, we take phase-space distributions of \barc pairs and 
evolve them according to the Fokker-Planck/Langevin equations, 
which describe nonequilibrium evolution of  phase-space distributions
during the QGP era. After it is finished and the medium
returns to the hadronic phase, our classical distribution
$f(\vec x, \vec p, t)$ has to be projected
into charmonium quantum states. 

He we examine how classical and quantum dynamics
correspond to each other in equilibrium, when both
are easily available. We simplify by
 examining the thermal distributions for a Coulombic system,
with $V\sim 1/r$. One can 
calculate the density of states for the classical system:
\begin{eqnarray}
Z & = & \int dr dp (4\pi)^2 r^2 p^2 \exp(-(p^2/2\mu-e^2/r)/T) \nonumber \\
  & = & \int dE \int dr dp (4\pi)^2 r^2 p^2 \exp(-E/T) \delta(E-p^2/2\mu
+1/r) \nonumber \\
  & = & \int dE \exp(-E/T) \int^{\infty}_{max(0,\sqrt{2\mu E}} dp 
\frac{e^2p^2}{(p^2/2\mu-E)^4} \nonumber
\end{eqnarray}
As one can see,
 this integral  is divergent for 
$E>0$. This means that 
this distribution is not normalizable and in thermal equilibrium all pairs 
are ionized. For $E<0$, we see from examining the partition function
\begin{equation}
\rho(E)\propto (-E)^{-5/2}
\end{equation}
Now we calculate the quantum mechanical density of states for $E<0$, 
\begin{equation}
\rho(E)=\Sigma_{n=1}^{\infty}n^2\delta(E+\frac{C}{n^2})
\end{equation}
which can be approximated by considering an integral:
\begin{eqnarray}
\int_{-\infty}^EdE^{'}\rho(E^{'}) & = & \Sigma_{i=1}^{\infty}i^2\theta(E+
\frac{C}{i^2}) \nonumber \\
 & \sim & E^{-3/2}\;{\rm for\;large\;enough\;E} \nonumber
\end{eqnarray}
Thus, $\rho(E) \sim E^{-5/2}$ for $E$ close to zero:  therefore for 
highly excited states the correspondence principle holds and the 
classical thermal distribution is recovered.
However, classical density of states is not 
so good of an approximation to the quantum-mechanical density of states
for the lowest bound state.

We also tested whether Langevin simulations leads to
the correct classical density of states, after some relaxation time.
The result of evolving an ensemble of \barc pairs at a fixed 
temperature according to classical Langevin dynamics
shown in Fig.\ref{etot_coulomb_fit}
 shows that equilibrium is indeed obtained.

\begin{figure}
\includegraphics[width=8cm]{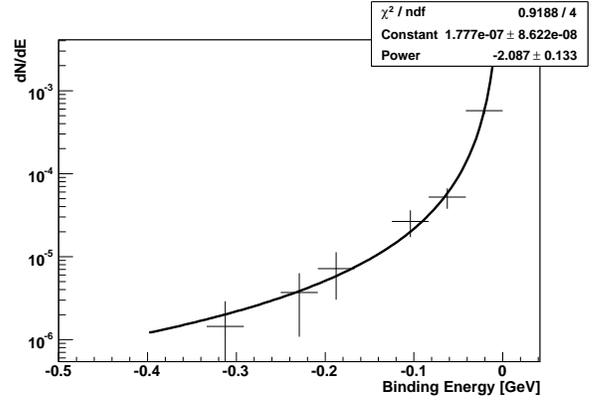}
\caption{The points show the density of states for 
 an ensemble of \barc pairs at a fixed 
temperature $T=1.5T_c$ obtained by long-time Langevin evolution,
compared to a fit with power 2.1 (curve), close  to classical 
nonrelativistic thermal distribution. }
\label{etot_coulomb_fit}
\end{figure}

\section{Coalescence Probability}

\label{prob}

After all distributions at the end of the QGP era are determined,
the next step is to calculate the probabilities of
it materializing as a particular charmonium state.

One approach is based on the Wigner probability distribution.
For any wavefunction $\psi (\pstn)$,
\be W_{\psi}(\pstn,\mmntm)= 
\frac{1}{\pi^3} \int d^3y \psi^* (\pstn+ \frac {\textbf{y}}{2}) \psi (\pstn- 
\frac{\textbf{y}}{2})e^{i\mmntm \cdot \textbf{y}} \ee
The wave function for J/$\psi$ was easily calculated numerically 
and then fitted to a
Gaussian $\psi_i(x)\propto e^{-3.8r^2/\fm^2}$. This leads immediately 
to the following parametrization of the
J/$\psi$'s Wigner distribution's (here  the constant is in 
$\fm^{-2}$) : 
\be \label{eqn_Wpsi}
W_{J/\psi}(x,p) = \frac{1}{\pi^3}exp(-7.6r^2-p^2/7.6)\ee

Many properties for the Wigner transform so defined may be discussed but for 
our purposes here we should note that
\be \int W_{\psi}(x,p)dp = |\psi (x)|^2 \ee
and for another wavefunction $\chi (x)$'s Wigner transform $W_{\chi}(x,p)$
\be \int W_{\psi}(x,p)W_{\chi}(x,p)dx dy
&=\frac{1}{(2\pi)^3}|\langle \psi |\chi \rangle|^2, \ee
because this is what we need to properly normalize our distributions and to 
compute overlaps. 

So, for our phase-space distribution at any given time $f(\vec x,\vec p,
\tau)$, we model our probability of pair being measured in the 
$J/\psi$-state as 
\be P_{J/\psi}(\tau)=(2\pi)^3 \int f(\pstn, \mmntm, \tau)
W_{J/\psi}(\pstn, \mmntm)d^3xd^3p \ee
We project the pairs onto the $J/\psi$-state not only at the onset
of the calculation, but also throughout the time evolution, monitoring
in this way the probability of $J/\psi$ production.
We use this approach to estimate the coalescence probability for the 
distribution in Section \ref{charmthermal}. 

Later on we switched to the ``energy projection method'', 
which is ultimately used for projection into $\chi,\psi^{'}$ states
as well as into \Jp.
It is based on the distribution over \barc energy, in the
pair rest frame, calculated with the (zero temperature) Cornell 
potential. One projection, used in Fig.\ref{raa_vs_npart},
is to all bound states, defined by
\begin{equation}
E_{CM}=V_{Cornell}(r)+\frac{p_{rel}^2}{M_c}<0.88\;\GeV{\rm .}
\end{equation}

Later on, we differentiate between various charmonium states again by 
examining the \barc pair's energy in its rest frame, and comparing it with 
the energies of solutions to Schr\"odinger's equation using the Cornell 
potential. For example, the lowest s-state solution, using the charm quark 
mass, has energy $E=0.33\;\GeV$, therefore we count all \barc pairs in 
our simulation with energy below $0.33\;\GeV$ as existing in the \Jp state.

\end{narrowtext}

\end{document}